**Discovery of highly-polarizable semiconductors BaZrS$_3$ and Ba$_3$Zr$_2$S$_7$**

Stephen Filippone[1], Boyang Zhao[2], Shanyuan Niu[2], Nathan Z. Koocher[3], Daniel Silevitch[4], Ignasi Fina[5], James M. Rondinelli[3], Jayakanth Ravichandran[2], R. Jaramillo[1†]

1. Department of Materials Science and Engineering, Massachusetts Institute of Technology, Cambridge, MA 02139, USA

2. Mork Family Department of Chemical Engineering and Materials Science, University of Southern California, Los Angeles, California 90089, USA

3. Department of Materials Science and Engineering, Northwestern University, Evanston, IL 60208, USA

4. Division of Physics, Math, and Astronomy, California Institute of Technology, Pasadena, CA 91125, USA

5. Institute de Ciència de Materials de Barcelona, Barcelona, Spain

† email: rjaramil@mit.edu

**Abstract**

There are few known semiconductors exhibiting both strong optical response and large dielectric polarizability. Inorganic materials with large dielectric polarizability tend to be wide-band gap complex oxides. Semiconductors with strong photoresponse to visible and infrared light tend to be weakly polarizable. Interesting exceptions to these trends are halide perovskites and phase-change chalcogenides. Here we introduce complex chalcogenides in the Ba-Zr-S system in perovskite and Ruddlesden-Popper structures as a new family of highly polarizable semiconductors. We report the results of impedance spectroscopy on single crystals that establish BaZrS$_3$ and Ba$_3$Zr$_2$S$_7$ as semiconductors with low-frequency relative dielectric constant ($\varepsilon_0$) in the range 50 – 100, and band gap in the range 1.3 – 1.8 eV. Our electronic structure calculations indicate the enhanced dielectric response in perovskite BaZrS$_3$ versus Ruddlesden-Popper Ba$_3$Zr$_2$S$_7$ is primarily due to enhanced IR mode-effective charges, and variations in phonon frequencies along ⟨001⟩; differences in the Born effective charges and the lattice stiffness are of secondary importance. This combination of covalent bonding in crystal structures more common to complex oxides results in a sizable Fröhlich coupling constant, which suggests that charge carriers are large polarons.

**Text**

Dielectric response controls the fundamental optoelectronic properties and the functional usefulness of semiconductors. The low-frequency dielectric response controls band bending and charge transport at semiconductor junctions, and the high-frequency response controls light-matter interaction. Dielectric response is also responsible for screening fixed and mobile charges, leading to polaron effects and controlling the interaction between charge carriers and lattice defects. In a recent contribution, we highlighted the relative dearth of semiconductors that combine band gap in the visible (VIS) and near-infrared (NIR) – the essential range for much device technology – with strong dielectric response [1]. We predicted that little-studied complex-structured chalcogenide materials may occupy this sparse region of the otherwise-well populated Ashby plot of dielectric constant *vs*. band gap.

Here we report experimental measurements and theoretical calculations of the dielectric response of two complex-structured chalcogenide semiconductors, $BaZrS_3$ in the perovskite structure, and $Ba_3Zr_2S_7$ in the Ruddlesden-Popper structure. We find that both are highly-polarizable, with low-frequency relative dielectric constant ($\varepsilon_{r,0}$) in the range of 50 – 100. Our results are consistent with expectations that semiconductors in the perovskite and related crystal structures are highly-polarizable, and they frame a hypothesis that polaron transport effects will be strong in chalcogenide perovskites. The strong low-frequency dielectric response is intriguing in light of the long radiative lifetime previously reported in $Ba_3Zr_2S_7$ [2].

$BaZrS_3$ is isostructural with $GdFeO_3$ in the distorted-perovskite structure, in space group *Pnma* [3]. It has a direct band gap of $E_g = 1.83$ eV, and strong optical absorption above the band gap [3,4]. $Ba_3Zr_2S_7$ forms in a two-layer Ruddlesden-Popper (RP) structure, in space group *P4$_2$/mnm* [2]. It has an indirect band gap at 1.25 eV, and a direct transition at 1.35 eV that results in strong optical absorption [2]. Both structures feature corner-sharing $ZrS_6$ octahedra, and the Zr-S-Zr bond geometry and covalence control the low-energy (*i.e.* near-band edge) electronic structure [5]. Both structures are non-polar, and $BaZrS_3$ has antiferroelectric cation order, illustrated in **Fig. 1**. $Ba_3Zr_2S_7$ is theoretically predicted to be close in energy to a polar, ferroelectric phase, with space group *Cmc*$2_1$ [6]. The dielectric properties of chalcogenides in the perovskite and RP structures have not been reported, beyond an early theoretical prediction of $\varepsilon_{r,0} = 46$ for $BaZrS_3$ [7].

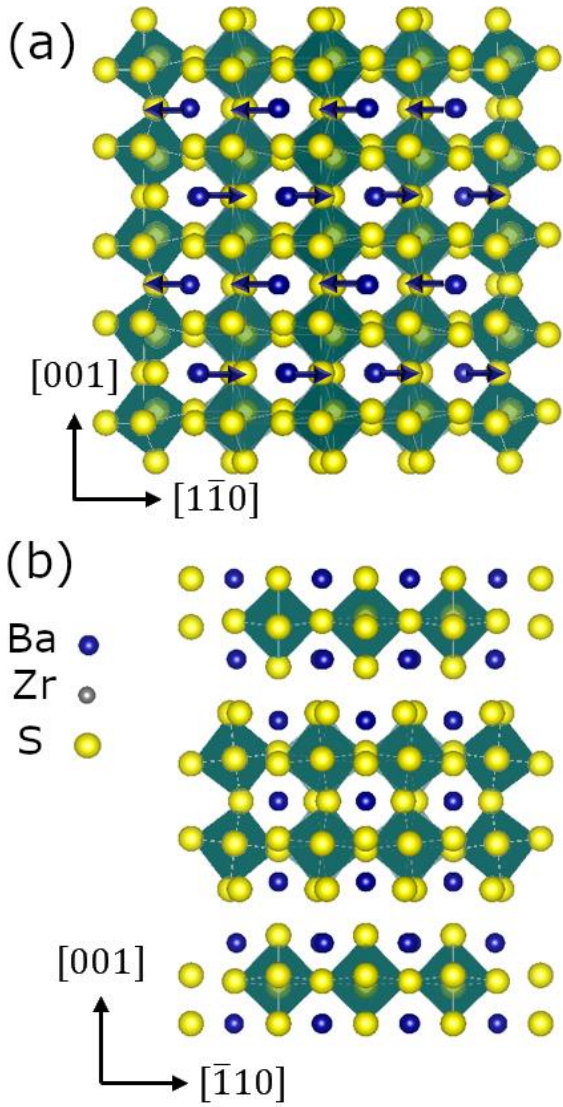

**Figure 1:** Crystal structures of (a) BaZrS$_3$ and (b) Ba$_3$Zr$_2$S$_7$. The blue, grey and yellow spheres represent Ba, Zr, and S respectively. ZrS$_6$ octahedra are shaded in green. The blue arrows indicate the antiferroelectric pattern of Ba$^{2+}$ cation displacement in BaZrS$_3$.

We performed impedance spectroscopy measurements on single-crystal samples that we grew using the salt flux growth technique, as reported previously [8]. We embedded the crystals in acrylic epoxy (SamplKwick, Buehler) to assist with sample handling and contacting. We ground and polished the epoxy handle to expose two opposite crystal faces. We formed parallel-plate capacitors by bonding Pt wire of diameter 0.001 in. to the two opposite crystal faces, using conductive silver epoxy (LCA-24, Bacon Adhesives). For some samples, we deposited 60 nm of Au by sputtering on the crystal faces before contacting the Pt wire. We made accurate measurements of the capacitor dimensions using micro X-ray computed tomography (μCT, X-Tek), as shown in **Fig. 2a-b**.

The small sample size presented a challenge to accurate impedance measurements. The typical crystal dimensions are $100 \times 100 \times 100$ $\mu m^3$, and the typical capacitance is in the range of 0.1 - 1 pF, making it essential to remove stray capacitance from the measurement circuit. To do so, we made custom test fixtures that maintain shielding to within ≈ 5 mm of the wired sample (**Fig. 2c**). With these custom fixtures, the stray capacitance of our measurement circuit without a sample was below the detection limit of our impedance analyzer. We performed temperature-dependent measurements between 2 – 300 K using a helium cryostat (PPMS DynaCool, Quantum Design) equipped with a coaxial-cabling insert.

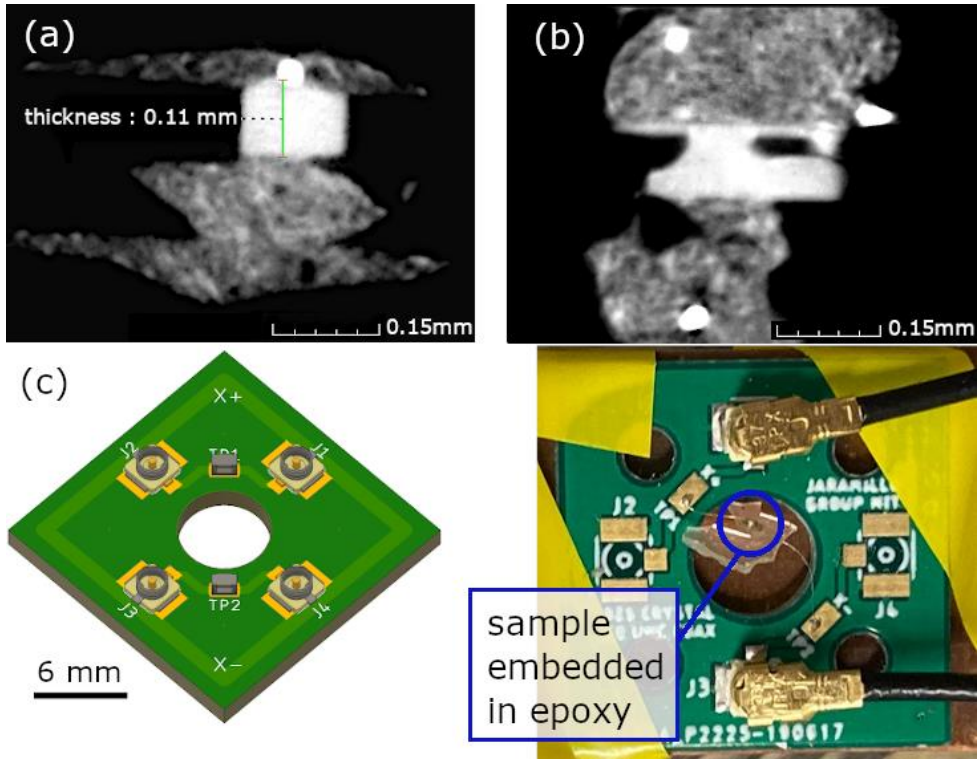

**Figure 2:** Illustration of techniques used to enable accurate impedance measurements on microscopic crystals. (a-b) $\mu$CT cross-section images used to determine the sample dimensions. Images show crystals and Ag epoxy contacts. Pt wires used for electrical leads appear as small, bright spots. (a) shows a BaZrS$_3$ crystal, with the thickness determined to be 0.11 mm. (b) shows a BaZrS$_3$ crystal for which $\mu$CT revealed a macroscopic void beneath the crystal surface; this sample was subsequently not used for impedance measurements. (c) Drawing and photograph of the custom test fixture used to enable measurements of sub-pF sample capacitance. The surface mounts are coax style UMC. The photograph shows a sample mounted in epoxy and wired to the fixture (upper-left and lower-right contact pads).

For temperature-dependent measurements we used an impedance analyzer (Hewlett Packard 4192A) in parallel R-C mode, 1 V AC drive, with zero adjustment over frequency range 1 kHz – 1 MHz. We also performed impedance spectroscopy measurements at in ambient conditions over a wider frequency range, from 0.1Hz to 1MHz, using a separate impedance analyzer (Novocontrol

Alpha-A, with Pot/Gal 30V-2A Test Interface). Our low-frequency measurements in ambient conditions suffered from a slow drift in sample resistance, possibly due to conduction through adsorbed water on the surface of the highly-resistive samples. To circumvent this problem, we used a shunt resistor to draw current at low frequency. This intervention made it possible to fit the Nyquist plots, and the sample impedance determined in this way is consistent with that measured directly at high frequency.

We performed density functional theory (DFT) calculations using the Vienna ab-initio simulation program with PAW pseudopotentials and the PBEsol functional [9–11]. The valence electronic configurations for the pseudopotentials are as follows: $5s^2 5p^4 6s^2$ for Ba, $4s^2 4p^6 5s^2 4d^2$ for Zr, $3s^2 3p^4$ for S. We employed a cut-off energy of 600 eV for both $BaZrS_3$ and $Ba_3Zr_2S_7$, as well as Γ-centered k-point meshes of $6 \times 6 \times 4$ ($BaZrS_3$) and $6 \times 6 \times 2$ ($Ba_3Zr_2S_7$). The electronic convergence threshold was $1 \times 10^{-8}$ eV and ionic force threshold was $1 \times 10^{-5}$ eV/Å for both structures. After relaxation, the stress tensor components for both structures were less than $8.6 \times 10^{-4}$ kbar. A gaussian smearing with 0.1 eV smearing width was also used for both structures. Dielectric properties were computed using density functional perturbation theory (DFPT).

For $BaZrS_3$ we find experimentally that $\varepsilon_{r,0} = 80 \pm 14$ representing seven unique samples and eleven independent measurements. For $Ba_3Zr_2S_7$ we find $\varepsilon_{r,0} = 72 \pm 37$, representing five unique samples and nine independent measurements. In **Fig. 3** we present detailed impedance analysis results for one representative sample each of $BaZrS_3$ and $Ba_3Zr_2S_7$. In **Figs. 3a** and **3d** we present the real component of the complex relative dielectric permittivity ($\varepsilon_r$) over the frequency range 1 kHz – 1 MHz between 2 – 300 K. The response for both samples is nearly constant with frequency and temperature, except for a rise at low frequency and high temperature, which is more pronounced for $BaZrS_3$. We take the response above $10^5$ Hz to be representative of the static dielectric constant.

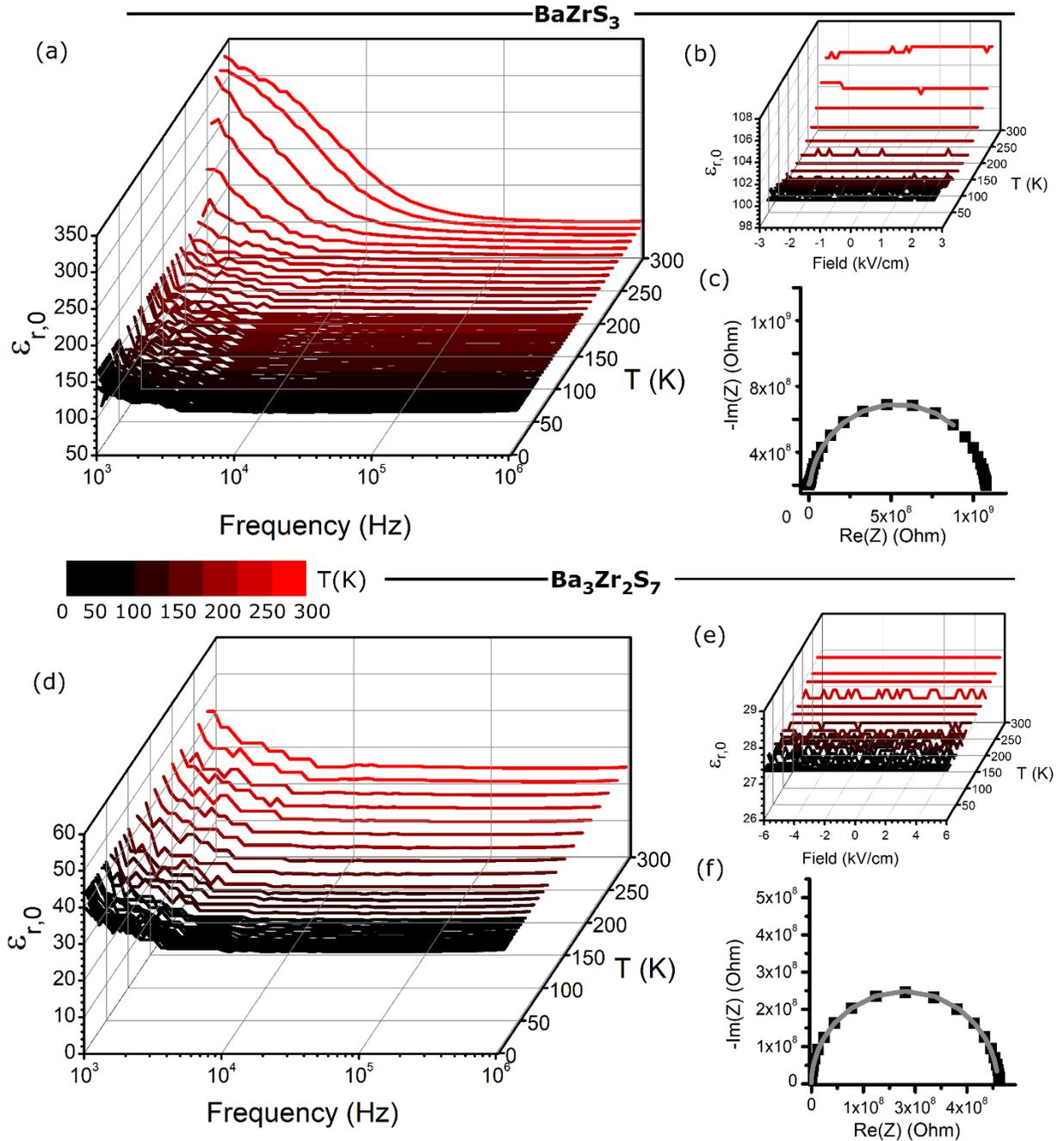

**Figure 3:** Impedance measurement results for one representative sample each of $BaZrS_3$ (a-c) and $Ba_3Zr_2S_7$ (d-f). (a, d) $\varepsilon_{r,0}$ measured for frequency between 1 kHz – 1 MHz and temperature between 2 – 300 K. (b, e) Dependence of $\varepsilon_{r,0}$ on an applied electric field. (c, f) Nyquist plot of complex impedance $(Z' - iZ'')$ measured between 0.1 Hz – 1 MHz ambient conditions; points are the data, and line is the fit.

In **Figs. 3b and 3e** we present the dependence of the dielectric response on a static electric field, which is added to the 1 V AC drive, measured at 100 kHz. We find no change in permittivity for electric field strength up to 2.7 kV/cm and 6 kV/cm for BaZrS$_3$ and Ba$_3$Zr$_2$S$_7$, respectively, at any temperature between 2 – 300 K. Ba$_3$Zr$_2$S$_7$ is paraelectric, but it is close in energy to a ferroelectric phase [6]. Therefore it may be expected to exhibit field-driven nonlinear capacitance, such as is observed in SrTiO$_3$ for field strength greater than 2 kV/cm at low temperature [12–14]. The crystals measured here were oriented with the electric field along the ⟨001⟩ crystal axis. We suggest that measurements with electric field applied along the ⟨110⟩ predicted polar distortion axis may reveal field-induced nonlinearity and phase transitions.

In **Figs. 3c** and **3f** we show representative Nyquist plots of the impedance data measured between 0.1Hz and 1MHz, as described above, together with fits using an R-CPE parallel circuit model. The results for sample impedance obtained from the Nyquist analysis is in good agreement with that obtained by direct capacitance measurements at high frequency.

Our experimental measurements of $\varepsilon_r$ are subject to a number of uncertainties. Calculating $\varepsilon_r$ from the measured capacitance relies on accurate measurements of the sample dimensions. Our µCT measurements are subject to a spatial resolution limit of 5 µm, which results in an uncertainty of $\approx$ 2% in $\varepsilon_{r,0}$. The statistical uncertainty in the best-fit parameters in the Nyquist analysis is $\approx$ 1%, and the statistical uncertainty in the high-frequency impedance measurements is lower still. We conclude that the spread in the dielectric constant (**Fig. 5**) comes from sample irregularities, including wiring and crystal quality, instead of measurement error. Ba$_3$Zr$_2$S$_7$ crystals produced by our salt-flux method have a minority Ba$_4$Zr$_3$S$_{10}$ phase intermixed, with the basal planes of the majority and minority phases aligned; it is unknown how this might affect our results. The Ag epoxy contacts often overhang the crystal, which can lead to a systematic error because the acrylic epoxy becomes part of the measured capacitor. However, the dielectric constant of this epoxy is small ($<$ 5) [15]. Therefore, the sample dominates the capacitance. All measured samples are highly-resistive (we estimate resistivity $\rho >$ 10MOhm for all samples), so that interface effects are likely negligible. We can rule out systematic errors due to porosity in the Ag epoxy contacts, because measurement made with sputtered gold contacts give quantitatively consistent results.

**Table 1**: Dielectric tensors for BaZrS$_3$ (**pink, bold**) and Ba$_3$Zr$_2$S$_7$ (*black, italic*).

| $\varepsilon_{ij}$ | $\varepsilon^{\text{ion}}$ | $\varepsilon^{\text{elec}}$ | $\bar{\varepsilon} = \dfrac{(\varepsilon_{11} + \varepsilon_{22} + \varepsilon_{33})}{3}$ |
|---|---|---|---|
| 11 | **68.6**, *44.5* | **10.3**, *9.0* | |
| 22 | **55.6**, *44.5* | **10.6**, *9.0* | **76.2**, *43.0* |
| 33 | **73.5**, *14.6* | **10.6**, *7.5* | |

To understand the atomic origins of the dielectric response, we computed the static dielectric tensor for BaZrS$_3$ and Ba$_3$Zr$_2$S$_7$ using DFT (**Table 1**). Our calculated average dielectric constant for BaZrS$_3$ is $\bar{\varepsilon} = 76.2$, including ionic and electronic contributions, and is in excellent agreement with the experimentally-measured value of $80 \pm 14$. We note that the dielectric response of BaZrS$_3$ was previously computed at the DFT-LDA level to be 46 [7]. Our analysis indicates that the discrepancy between calculated values is due to lower vibrational mode frequencies in our calculations [7]. Our calculated average dielectric constant for $Ba_3Zr_2S_7$ is $\bar{\varepsilon} = 43.0$, which is within the uncertainty of the measured value of $72 \pm 37$, and substantially smaller than for perovskite. The experimentally-measured values are nearly constant with frequency (1 kHz – 1MHz) and temperature (2 – 300 K), meaning that comparison with zero temperature DFT results is meaningful. The electronic contribution to the dielectric response ($\varepsilon^{elec}$) is similar for $BaZrS_3$ and $Ba_3Zr_2S_7$, although slightly larger for perovskite. This is consistent with our understanding that the electronic structures of the two phases are similar, consisting of filled sulfide $3p$ states at the valence band edge, and unoccupied Zr $4d$ states at the conduction band edge. [5] This leads to similar changes in orbital hybridization (and covalency) under applied electric fields. Therefore, the differences in total dielectric response originate from difference in the ionic response by the phonons in the two materials.

The contribution of each optically-active phonon mode $\mu$ to the total ionic dielectric response, here averaged over the principal axes, is given as

$$\varepsilon^{ion} = \sum_\mu \left( \frac{1}{3} \sum_\alpha \frac{1}{4\pi^2 \varepsilon_0 V} \frac{(Z^*_{\mu\alpha})^2}{\nu_\mu^2} \right), \qquad (1)$$

where $Z^*_{\mu\alpha}$ are the effective charges for mode $\mu$, obtained as a mass-normalized sum of the atomic Born effective charges tensors (available in Supplementary Table S2 and S3) for each infrared (IR-active) eigenvector [7]. The frequency-squared of each IR mode $\nu_\mu^2$ appears in the denominator of the summation in **Eq. 1**. These individual mode contributions add together linearly to obtain $\varepsilon^{ion}$; therefore, a large ionic contribution can be due to large mode-effective charges, low-frequency IR-active phonons, or an increase in number of modes owing to the complexity of the asymmetric unit in the unit cell.

The perovskite $BaZrS_3$ possesses 57 optical modes, of which 25 ($\Gamma_{IR} = 9B_{1u} + 7B_{2u} + 9B_{3u}$) are IR-active. $Ba_3Zr_2S_7$ possesses 141 optical modes, of which 50 are IR-active ($\Gamma_{IR} = 10A_{2u} + 20E_u$). Thus, the difference in connectivity of the octahedral layers between the two structures activates more oscillators in the RP phase; however, these IR modes can contribute in different ways to $\varepsilon^{ion}$. Furthermore, although the atomic Born effective charges for these two sulfides are similar, the displacement patterns (and hence relative motion of ions) can be significantly different owing to octahedral-connectivity disruption. The origin of the reduced ionic dielectric response in $Ba_3Zr_2S_7$ then is due either to stiffer vibrational modes, or smaller mode-effective charges. Our analysis of the low-frequency phonon densities-of-states of the two phases indicates that the difference in the vibrational frequencies between the two phases are minor for the in-plane directions (Figure S1 in the SI), but are more pronounced in the out-of-plane directions, due to a stiffening of the lattice and higher phonon frequencies along the [001] direction of $Ba_3Zr_2S_7$.

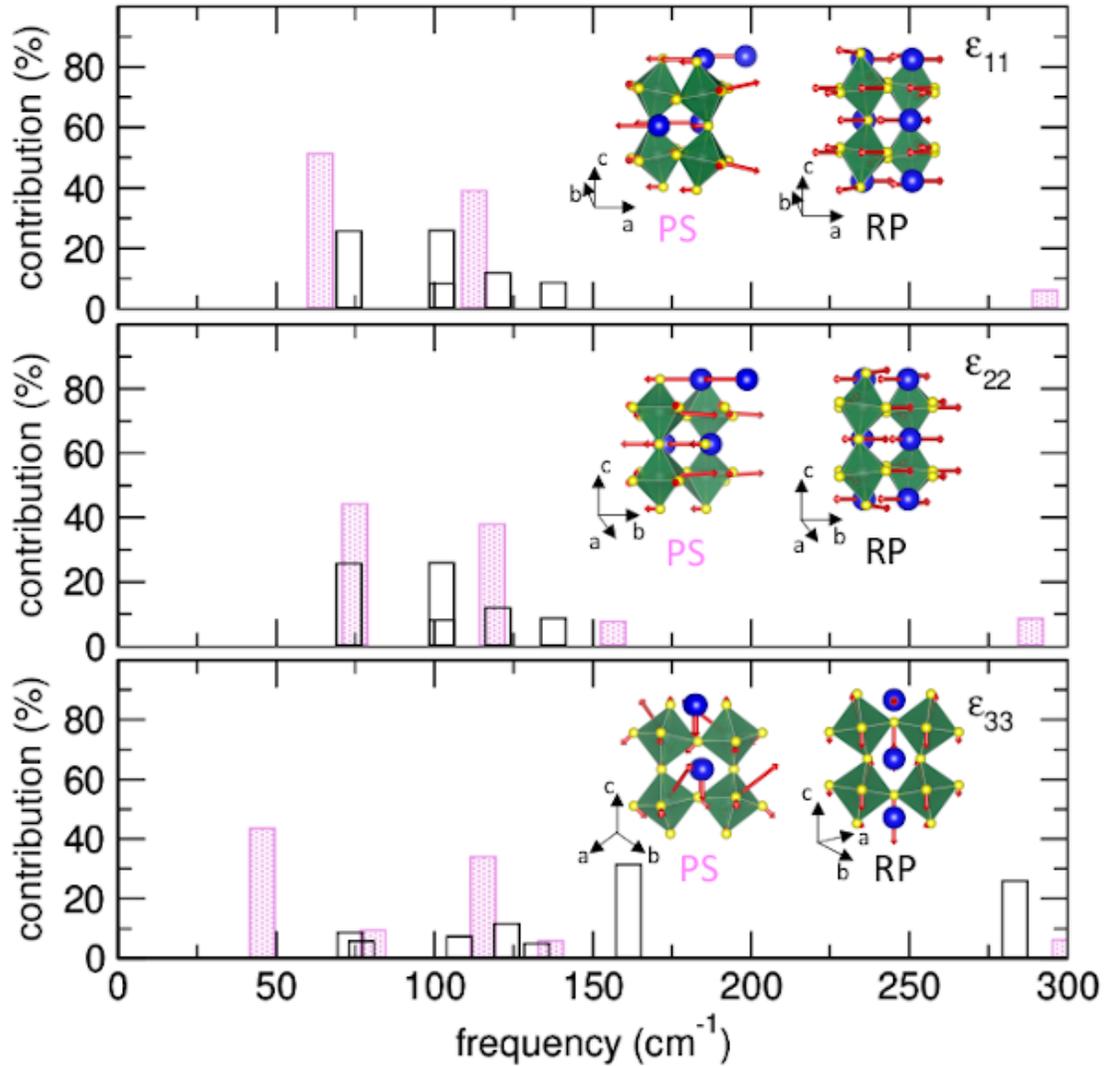

**Figure 4**: Phonon mode contributions to the ionic part of the low-frequency dielectric tensor for $BaZrS_3$ (pink) and $Ba_3Zr_2S_7$ (black). The abscissa indicates the phonon frequencies; modes that contribute less than 5% of the total dielectric response are omitted for clarity. The insets show atomic motions for the dominant mode in each direction for each material: for $\varepsilon_{11}$, 63.8 cm$^{-1}$ for PS, 102.2 cm$^{-1}$ for RP; for $\varepsilon_{22}$, 74.8 cm$^{-1}$ for PS, 102.2 cm$^{-1}$ for RP; for $\varepsilon_{33}$, 45.6 cm$^{-1}$ for PS, 161.3 cm$^{-1}$ for RP. Atomic displacements are indicated with arrows; PS = perovskite.

**Fig. 4** compares the relative contribution of each IR mode to the ionic dielectric response along each principal axis. For $\varepsilon_{11}$ we find that BaZrS$_3$ exhibits two dominant IR modes, contributing at 63.8 cm$^{-1}$ ($B_{1u}$ symmetry) and 112.5 cm$^{-1}$ ($B_{1u}$), and Ba$_3$Zr$_2$S$_7$ also has two dominant modes at similar frequencies (73.0 and 102.2 cm$^{-1}$, both of $E_u$ symmetry); we also find that the mode-effective charges are similar for these four modes. The BaZrS$_3$ mode at $\nu = 63.8$ cm$^{-1}$ is the lowest-frequency among these four, and makes the largest contribution to the ionic response. It is comprised predominately of Ba, Zr and apical S displacements along the $x$ direction, whereas the BaZrS$_3$ mode at $\nu = 112.5$ cm$^{-1}$ includes motions of axial and equatorial S and Ba along the $x$ direction. The RP mode at 73.0 cm$^{-1}$ ($E_u$) has Zr, equatorial S, interfacial Ba, and slight axial S displacements along the $x$ direction. The RP mode at 102.2 cm$^{-1}$ ($E_u$) has motion of Ba, equatorial S, and axial S along the $x$ direction (see Supporting Information for videos of the vibrational modes). We attribute the lower frequency of the dominant vibrational mode in the perovskite, compared to RP, to the shorter Zr-S bond distances for the apical sites in Ba$_3$Zr$_2$S$_7$, leading to mode stiffening. Interestingly, although the perovskite phase is denser than the RP phase (24.6 Å$^3$/atom vs. 26.3 Å$^3$/atom, respectively, at the DFT level), its vibrational modes are softer. We reconcile this discrepancy by comparing the RP bilayer volume-per-ZrS$_6$ octahedron to the perovskite, to eliminate contributions from BaS rock-salt layer from the analysis. With the rock-salt layers omitted, we find that the RP has a volume of 228.0 Å$^3$ per octahedron, compared to 245.8 Å$^3$ per octahedron for the perovskite, thereby justifying the softer phonon modes in the perovskite. This comparison is justified because the active in-plane phonons in Ba$_3$Zr$_2$S$_7$ derive from motions within the octahedral bilayer (see SI).

The response and explanation for $\varepsilon_{22}$ is similar to that for $\varepsilon_{11}$. In the case of $\varepsilon_{22}$ the dominant modes for the perovskite and the RP are closer in frequency than for $\varepsilon_{11}$, and again have comparable mode-effective charges. Therefore, the difference between $\varepsilon_{22}$ for the two phases is smaller than for $\varepsilon_{11}$. The second-most dominant perovskite mode at 118.3 cm$^{-1}$ ($B_{3u}$) is higher in frequency than the second-most dominant RP mode (102.2 cm$^{-1}$, $E_u$ symmetry). This suggests that the $a$ axis of the perovskite is mechanically softer than the $b$ axis; indeed, $a/b$=1.02, and thus the vibrational modes along [100] in BaZrS$_3$ are lower in frequency than those along [010]. Similar to the behavior along the $x$ direction, the 74.8 cm$^{-1}$ ($B_{3u}$) perovskite mode has motion of Zr, apical S, and Ba along the $y$ direction, while the 118.3 cm$^{-1}$ mode has motion of all S atoms and Ba along the $y$ direction. The 73.0 cm$^{-1}$ ($E_u$) RP mode has motion of Zr, equatorial and axial S, and interfacial Ba along the $y$ direction, while the 102.2 cm$^{-1}$ ($E_u$) mode has motion of Ba and equatorial and axial S in the $y$ direction.

The out-of-plane response $\varepsilon_{33}$ shows the largest contrast between the two phases, as expected due to the broken octahedral corner-connectivity along the long axis of Ba$_3$Zr$_2$S$_7$. We find two dominant modes for each phase, and the dominant perovskite mode at 45.6 cm$^{-1}$ ($B_{2u}$) occurs at lower frequency than the dominant RP mode at 161.3 cm$^{-1}$ ($A_{2u}$). This perovskite mode at 45.6 cm$^{-1}$ involves motions of all atoms along the $z$ direction. The RP mode at 161.3 cm$^{-1}$ also involves all atoms moving along the $z$ direction, with the interfacial Ba and S atoms exhibiting larger displacements than the atoms in the bilayer. There are several additional modes in the range 60-150 cm$^{-1}$ that contribute to $\varepsilon_{33}$ in the RP.

The different $\varepsilon_{33}$ response between the two phases can be further understood by analyzing the mode effective charges (MECs). The dominant, low frequency mode in the perovskite (45.6 cm$^{-1}$, $B_{2u}$) has a MEC that is approximately 100% larger than the low-frequency modes in the RP, which make relatively little contribution to the ionic response: -0.815 $e$Å for the perovskite mode at 45.6 cm$^{-1}$ ($B_{2u}$) vs. -0.418 $e$Å and -0.357 $e$Å for the RP modes at 73.2 cm$^{-1}$ ($A_{2u}$) and 76.9 cm$^{-1}$ ($A_{2u}$), respectively. The second-most dominant mode in the perovskite, at 115.3 cm$^{-1}$ ($B_{2u}$), has a MEC of -1.82 $e$Å. For the RP, the dominant mode at 161.3 cm$^{-1}$ ($A_{2u}$) has a MEC of -1.75 $e$Å, and the second-most dominant mode at 283.4 cm$^{-1}$ ($A_{2u}$) has a MEC of -2.79 $e$Å. The contribution of these large MEC values to the ionic response is curtailed by the higher mode frequencies (c.f. the $v_\mu^2$ term in the denominator of Eq. (1)), with the net effect that the out-of-plane ionic response of Ba$_3$Zr$_2$S$_7$ is substantially suppressed.

BaZrS$_3$ and Ba$_3$Zr$_2$S$_7$ have among the strongest low-frequency dielectric response reported for semiconductors with band gap in the VIS-NIR. In **Fig. 5** we summarize our results by presenting our measured values on an Ashby plot of low-frequency dielectric constant ($\varepsilon_{r,0}$) vs. $E_g$, and including a number of well-known semiconductor materials. Materials based on light elements and a motif of tetrahedral covalent bonding, including many group-IV, III-V, and II-VI semiconductors, have $\varepsilon_{r,0} \approx 10$. Stronger low-frequency dielectric response is associated with more complex crystal structures, such as complex oxides with band gap in the ultraviolet, and lead salts with band gap in the mid-IR. Among semiconductors with band gap in the VIS-NIR, the strongest dielectric response is found among materials with complex crystal structures and mixed bonding motifs, such as layered SnS, halide perovskites, and the complex chalcogenides reported here.

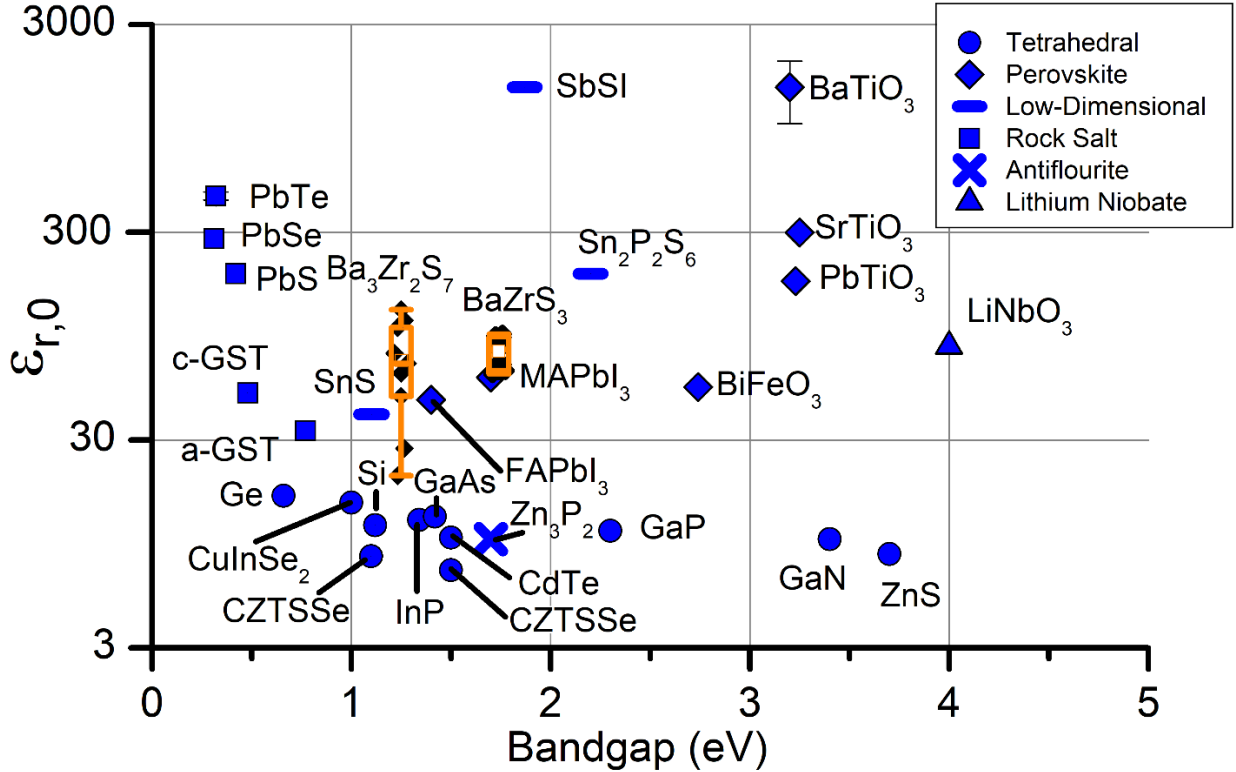

**Figure 5:** Ashby plot of low-frequency relative dielectric constant ($\varepsilon_{r,0}$) vs. band gap, showing our results for BaZrS$_3$ and Ba$_3$Zr$_2$S$_7$ together with results for a number of well-known semiconductor materials. The results for BaZrS$_3$ and Ba$_3$Zr$_2$S$_7$ are presented as red box-and-whisker symbols and black points, which include measurements on multiple samples and using multiple techniques: seven unique samples and eleven independent measurements for BaZrS$_3$, five unique samples and nine independent measurements for Ba$_3$Zr$_2$S$_7$. Data for all but the lead salts is as presented in ref. [1]; data for lead salts is as reported in ref. [16].

The large low-frequency dielectric response suggests that polaron effects may be prominent in the charge transport properties and optical spectra of these and related complex chalcogenide semiconductors. The Fröhlich coupling constant ($\alpha$) parameterizes the interaction between charge carriers and polar phonons, and the tendency toward polaron formation. We estimate $\alpha \approx 0.8 - 2$ for BaZrS$_3$, and $\alpha \approx 1$ for Ba$_3$Zr$_2$S$_7$, using available data for phonon frequency and electron effective mass [4,17,18]. These are comparable to the range reported for lead halide perovskites (1.5 − 2.7), and implies that charge carriers may be large polarons [19,20].

In conclusion, we find that complex chalcogenides in the Ba-Zr-S system in perovskite and Ruddlesden-Popper structures are highly-polarizable semiconductors, with low-frequency dielectric constant comparable to or even exceeding the highest reported values for semiconductors with band gap in the NIR-VIS, including halide perovskites. We showed that the anisotropy in the dielectric response is sensitive to the 3D connectivity of the transition metal-sulfide octahedra, which manifest in changes in IR mode-effective charges and frequencies. This family of complex chalcogenide semiconductors therefore combines strong optical absorption, excellent environmental stability, and strong dielectric response, and consists of abundant and non-toxic elements [3,4,18]. It is an interesting and open question whether the strong dielectric polarizability reported here is related to the long radiative lifetime reported previously in $Ba_3Zr_2S_7$, as suggested for halide perovskites [2,21–24].

**Acknowledgments**
We acknowledge support from the National Science Foundation (NSF) under grant no. 1751736, "CAREER: Fundamentals of Complex Chalcogenide Electronic Materials," from the MIT Skoltech Program, and from "la Caixa" Foundation MISTI Global Seed Funds Financial support from the Spanish Ministry of Economy, Competitiveness and Universities, through the "Severo Ochoa" Programme for Centres of Excellence in R&D (SEV-2015-0496) and the MAT2015-73839-JIN (MINECO/FEDER, EU) and MAT2017-85232-R (AEI/FEDER, EU) projects, and from Generalitat de Catalunya (2017 SGR 1377) is acknowledged. IF acknowledges Ramón y Cajal contract RYC-2017-22531. SF acknowledges support from the NSF Graduate Research Fellowship under grant no. 1122374. The work at Caltech was supported by National Science Foundation Grant No. DMR-1606858. JR, BZ and SN acknowledge support from Army Research Office under Award No. W911NF-19-1-0137 and Air Force Office of Scientific Research under Award No. FA9550- 16-1-0335. N.Z.K. and J.M.R. acknowledge support from the U.S. Department of Energy (DOE) under grant number DE-SC0012375 and the DOD-HPCMP for computational resources.  N.Z.K. thanks Dr. Michael Waters and Dr. Xuezeng Lu for helpful discussions.

**Supplemental Information**

**Table S1:** List of studied samples, their names as recorded in the laboratory records, and their contributions to the figures in the text.

| Composition | Sample Name | Contribution to Figures |
|---|---|---|
| $BaZrS_3$ | BZS113-190305 | Figure 5 |
| $BaZrS_3$ | BZS113-190306 | Figure 5 |
| $BaZrS_3$ | BZS113-190415 | Figure 5 |
| $BaZrS_3$ | BZS113-190419-1 | Figure 3, Figure 5 |
| $BaZrS_3$ | BZS113-190514-1 | Figure 5 |
| $BaZrS_3$ | BZS113-190514-2 | Figure 5 |
| $BaZrS_3$ | BZS113-190514-3 | Figure 5 |
| $Ba_3Zr_2S_7$ | BZS327-190508 | Figure 5 |
| $Ba_3Zr_2S_7$ | BZS327-190509 | Figure 3, Figure 5 |
| $Ba_3Zr_2S_7$ | BZS327-190510 | Figure 5 |
| $Ba_3Zr_2S_7$ | BZS327-190116-1 | Figure 5 |
| $Ba_3Zr_2S_7$ | BZS327-180831 | Figure 5 |

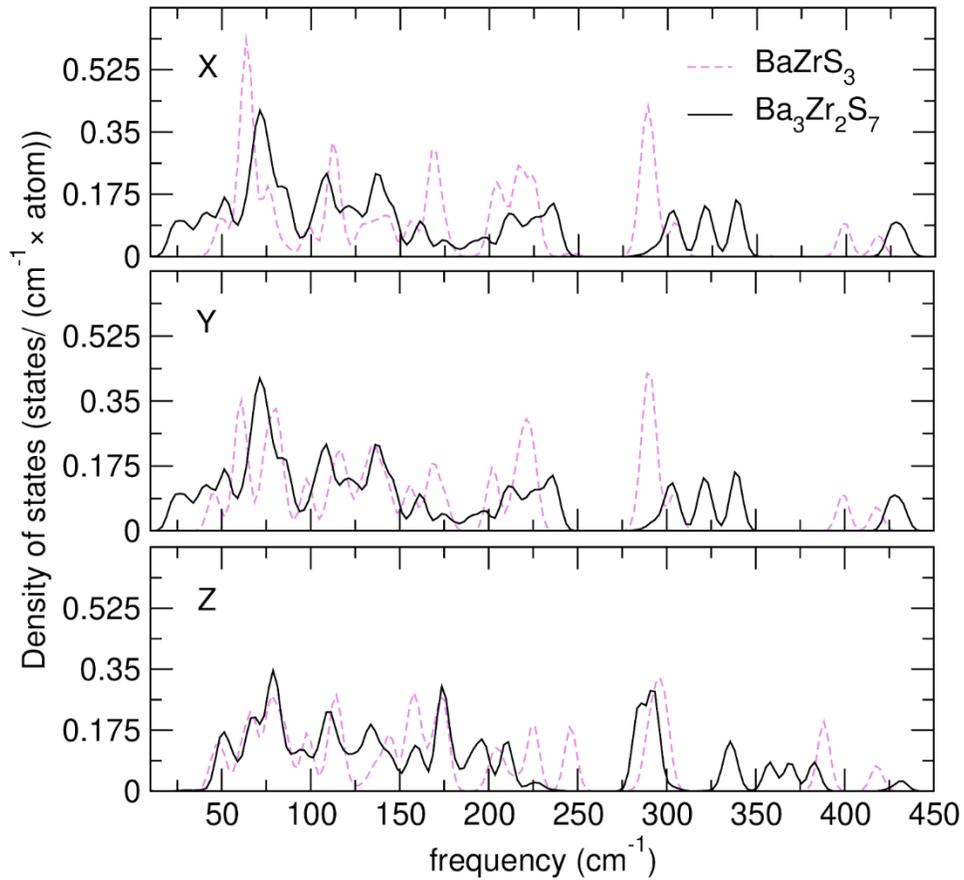

**Figure S1**: Direction-resolved phonon density of states for BaZrS$_3$ (pink) and Ba$_3$Zr$_2$S$_7$ (black), where the *a*, *b*, and *c* axes coincide with the X, Y, and Z directions, respectively. Ba$_3$Zr$_2$S$_7$ exhibits softer modes below 100 cm$^{-1}$ in the X and Y directions compared to BaZrS$_3$, but comparable mode frequencies beginning at 50 cm$^{-1}$ along the Z direction.

**Table S2**: Born effective charge tensors (*e*) for each unique atomic site in BaZrS$_3$. The Wyckoff orbits for each site are shown next to the atom label, and the corresponding tensor is shown below.

| Ba (4c) | | |
|---|---|---|
| 2.71 | -0.233 | 0.00 |
| -0.208 | 2.67 | 0.00 |
| 0 | 0 | 2.69 |

| Zr (4b) | | |
|---|---|---|
| 7.86 | -0.929 | -1.14 |
| 0.723 | 8.08 | -2.27x10$^{-3}$ |
| 0.865 | -3.43x10$^{-2}$ | 8.23 |

| S1 (4c) | | |
|---|---|---|
| -2.27 | 2.10x10$^{-2}$ | 0.00 |
| -0.227 | -2.01 | 0.00 |
| 0 | 0 | -6.51 |

| S2 (8d) | | |
|---|---|---|
| -4.16 | -2.10 | -5.56x10$^{-2}$ |
| -2.14 | -4.38 | -4.57x10$^{-2}$ |
| 1.07x10$^{-2}$ | -7.69x10$^{-2}$ | -2.21 |

**Table S3**: Born effective charge tensors (*e*) for each unique atomic site in Ba$_3$Zr$_2$S$_7$. The Wyckoff orbits for each site are shown next to the atom label, and the corresponding tensor is shown below.

| Ba1 (8j) | | |
|---|---|---|
| 2.52 | -7.69x10$^{-2}$ | -7.27x10$^{-2}$ |
| -7.18x10$^{-2}$ | 2.52 | 7.84x10$^{-2}$ |
| 0.112 | -0.112 | 3.21 |
| | | |
| Ba2 (4*f*) | | |
| 2.92 | -9.16x10$^{-2}$ | 0.00 |
| -8.90x10$^{-2}$ | 2.92 | 0.00 |
| 0.00 | 0.00 | 2.15 |
| | | |
| Zr1 (8*j*) | | |
| 8.02 | -0.162 | -0.211 |
| -0.162 | 8.02 | -0.214 |
| 0.524 | 0.524 | 5.88 |
| | | |
| S1 (4*e*) | | |
| -4.24 | -1.69 | 0 |
| -1.69 | -4.24 | 0 |
| 0 | 0 | -1.67 |
| | | |
| S2 (4*g*) | | |
| -2.32 | 0.110 | 0 |
| 0.110 | -2.32 | 0 |
| 0 | 0 | -4.82 |
| | | |
| S3 (8*j*) | | |
| -2.23 | 1.35x10$^{-2}$ | -0.159 |
| 1.35x10$^{-2}$ | -2.23 | -0.158 |
| -3.51x10$^{-2}$ | -3.51x10$^{-2}$ | -4.64 |
| | | |
| S4 (4*e*) | | |
| -4.36 | -1.85 | 0 |
| -1.85 | -4.36 | 0 |
| 0 | 0 | -1.46 |
| | | |
| S5 (8h) | | |
| -4.36 | 1.98 | 0 |
| 1.95 | -4.29 | 0 |
| 0 | 0 | -1.57 |
| | | |

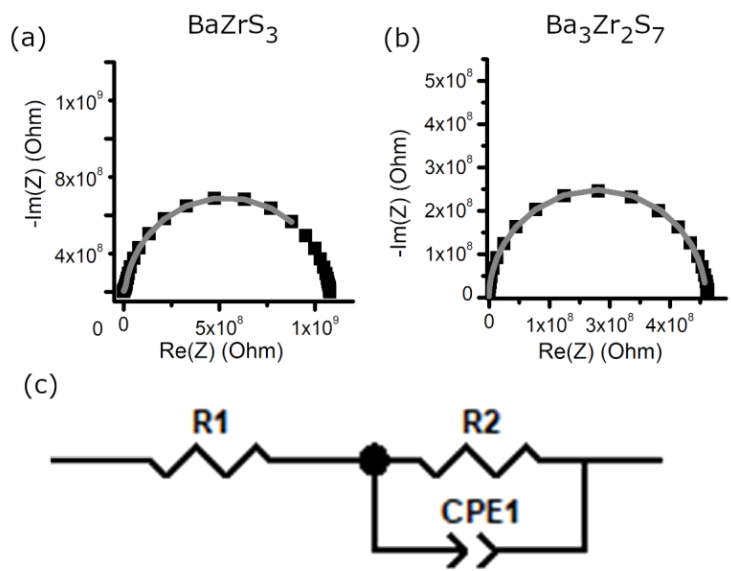

**Figure S2**: Impedance spectroscopy data, fits, and model. (a-b) reproduce the data and fits presented in Fig. 3 for $BaZrS_3$ (a, sample BZS113-190419-1) and $Ba_3Zr_2S_7$ (b, sample BZS327-190509). (c) Equivalent circuit model used to fit the data, using the software ZView (Scribner Associates, Inc.).; R1 and R2 are resistors, CPE1 is a constant-phase element.

**Table S4:** Table showing the parameters for the impedance spectroscopy fits presented in Fig. S2.

|  | BZS113-190419-1 | BZS327-190509 |
|---|---|---|
| Chi-Sqr | 5.1587E+12 | 2.7329E+14 |
| Sum-Sqr | 5.5714E+14 | 1.6944E+16 |
| R1($\pm$) | 3100900 | -10908000 |
| R1(Error) | 498950 | 3397700 |
| R1(Error%) | 16.09 | 31.149 |
| R2($\pm$) | 1074000000 | 9073700000 |
| R2(Error) | 697490 | 20095000 |
| R2(Error%) | 0.06494 | 0.22146 |
| CPE1-T($\pm$) | 1.1659E-12 | 4.1829E-13 |
| CPE1-T(Error) | 8.9966E-15 | 3.9183E-15 |
| CPE1-T(Error%) | 0.77164 | 0.93674 |
| CPE1-P($\pm$) | 0.94169 | 0.95677 |
| CPE1-P(Error) | 0.00109 | 0.00155 |
| CPE1-P(Error%) | 0.11624 | 0.16237 |